\documentclass[
    ,final            
  ,numberedheadings 
  ]
  {aipproc}

\layoutstyle{6x9}

\begin{document}

\title{The Effects of Changes in Reaction Rates on Simulations of Nova Explosions}



\author{S. Starrfield}{address={School of Earth and Space Exploration, Arizona
State University, Tempe, AZ 85287-1404:sumner.starrfield@asu.edu}
}

\author{C. Iliadis}{address={Department of Physics and Astronomy, University of North
Carolina, Chapel Hill, NC27599-3255:iliadis@unc.edu}}

\author{W. R. Hix}{address={Physics Division, Oak Ridge National Laboratory, Oak Ridge, TN
37831-6354:raph@ornl.gov}}

\author{F. X. Timmes}{address={X-2, Los Alamos National Laboratory, Los Alamos,
NM, 87545:fxt44@mac.com}}

\author{W. M. Sparks}{address={Science Applications
International Corporation, San Diego CA, 92121 \& X-4, Los Alamos
National Laboratory, Los Alamos, NM, 87545:wms@lanl.gov}}

\begin{abstract}
Classical novae participate in the cycle of Galactic chemical
evolution in which grains and metal enriched gas in their ejecta,
supplementing those of supernovae, AGB stars, and Wolf-Rayet
stars, are a source of heavy elements for the ISM.  Once in the
diffuse gas, this material is mixed with the existing gases and
then incorporated into young stars and planetary systems during
star formation. Infrared observations have confirmed the presence
of carbon, SiC, hydrocarbons, and oxygen-rich silicate grains in
nova ejecta, suggesting that some fraction of  the pre-solar
grains identified in meteoritic material come from novae. The mean
mass returned by a nova outburst to the ISM probably exceeds $\sim
2\times 10^{-4}$ M$_\odot$.  Using the observed nova rate of
35$\pm$11 per year in our Galaxy, it follows that novae introduce
more than $\sim 7\times 10^{-3}$ M$_\odot$ yr$^{-1}$ of processed
matter into the ISM.  Novae are expected to be the major source of
$^{15}$N and $^{17}$O in the Galaxy and to contribute to the
abundances of other isotopes in this atomic mass range. Here, we
report on how changes in the nuclear reaction rates affect the
properties of the outburst and alter the predictions of the
contributions of novae to Galactic chemical evolution.

\end{abstract}

\maketitle

\section{Introduction}

The observable consequences of accretion onto white dwarfs (WDs)
include the Classical (CN), Symbiotic, and Recurrent Nova (RN)
outbursts, and the possible evolution of the Super Soft, Close
Binary, X-ray Sources (SSS) to Type Ia Supernova (SN Ia)
explosions.  This diversity of phenomena occurs because of
differences in the properties of the secondary star, the mass of
the WD, and the stage of evolution of the binary system (the
orbital period, the luminosity of the WD and the rate of mass
accretion onto the WD). A CN explosion occurs in the accreted
hydrogen-rich envelope on the low-luminosity WD component of a
Cataclysmic Variable (CV) system. One dimensional (1D)
hydrodynamic studies, which follow the evolution of the material
falling onto the WD from a bare core to the explosion, show that
the envelope grows in mass until it reaches a temperature and
density at its base that is sufficiently high for ignition of the
hydrogen-rich fuel to occur. Both observations of the chemical
abundances in CN ejecta and theoretical studies of the
consequences of the thermonuclear runaway (TNR) in the WD envelope
strongly imply that mixing of the accreted matter with core matter
occurs at some time during the evolution to the peak of the
explosion. How and when the mixing occurs is not yet known (see,
e. g., Starrfield, Iliadis, and Hix 2006:S06 for a discussion).

If the bottom of the accreted layer is sufficiently degenerate and
well mixed with the core, then a TNR occurs and explosively ejects
core plus accreted material in a fast CN outburst.  The evolution
of nuclear burning on the WD, and the total amount of mass that it
accretes and ejects depends upon: the mass and luminosity of the
underlying WD, the rate of mass accretion onto the WD, the
chemical composition in the reacting layers (which includes the
metallicity of the CV system), the mixing history of the envelope,
and the outburst history of the system.

The high levels of enrichment of novae ejecta in elements ranging
from carbon to sulfur confirm that there is significant dredge-up
of matter from the core of the underlying WD and enable novae to
contribute to the chemical enrichment of the interstellar medium
(Gehrz et al. 1998: G98). Observations of the epoch of dust
formation in the expanding shells of novae allow important
constraints to be placed on the dust formation process and confirm
that graphite, SiC, and SiO$_2$ grains are formed by the outburst
(G98 and references therein).  It is possible that grains from
novae were injected into the pre-solar nebula and can be
identified with some of the pre-solar grains or ``stardust'' found
in meteorites (Zinner 1998, Amari et al. 2001, Jos\'e et al.
2004).  Finally, $\gamma$-ray observations during the first
several years of their outburst, done with the next generation of
satellite observatories, could confirm the presence of decays from
$^7$Be and $^{22}$Na (Weiss and Truran 1990; Nofar et al. 1991;
Jean et al. 2000, and references therein). In the next section we
report on NOVA our one-dimensional (1D) hydrodynamic computer code
that we have used for the new calculations done with the reaction
rate library of Iliadis (2005, private communication).  We follow
that with a discussion of our evolutionary results and the
implications of the new rates for the nova outburst. We end with a
summary.

\section{The Hydrodynamic Computer Code and Nuclear Reaction Rate
Libraries}

NOVA is a spherically symmetric , fully implicit, Lagrangian,
hydrodynamic computer code that incorporates a large nuclear
reaction rate network. It is described in detail in Starrfield et
al. (1998: S98), Starrfield et al. (2000:S00; and references
therein). As reported in those papers, we have found that
improving the opacities, equations-of-state, and the nuclear
reaction rates have had important effects on both the energetics
and the nucleosynthesis. Similar results have been found in the
calculations of the Barcelona group as reported elsewhere (Hernanz
and Jos\`e 2000, and references therein).  Therefore, over the
past few years we continued to improve the physics in NOVA and
then determined the effects of the improved physics on simulations
of the CN outburst (S06, and references therein).  A major effort
has been the effects of improving the reaction rates used in the
calculations on the evolution of the CN outburst and the resulting
nucleosynthesis.  In this paper we compare our earlier studies to
a recent reaction rate library of Iliadis (current as of August
2005). Since NOVA is always being updated and improved, for the
work to be reported on in this paper we have made one major change
and numerous minor changes.

The major change is that we no longer use the nuclear reaction
network of Weiss and Truran (1990: WT90) but have switched to the
modern nuclear reaction network of Hix and Thielemann (1999:
HT99).  While both networks utilize reaction rates in the common
REACLIB format and perform their temporal integration using the
Backward Euler method introduced by Arnett and Truran (1969:
AT69), there are two important differences.  First, WT90 implement
a single iteration, semi-implicit backward Euler scheme, which has
the advantage of a relatively small and predictable number of
matrix solutions, but allows only heuristic checks that the chosen
time step results in a stable or accurate solution. HT99 implement
the iterative, fully implicit backward Euler scheme, repeating the
Backward Euler step until convergence is achieved, providing a
measure of the stability and accuracy. If convergence does not
occur within a reasonable number of iterations, the time step is
subdivided into smaller intervals until a converged solution is
achieved. This allows the fully implicit backward Euler
integration to respond to instability or inaccuracy in a way that
is impossible with the semi-implicit backward Euler approach. As a
result, the fully iterative approach can often safely employ
larger time steps than the semi-implicit approach, obviating the
speed advantage of the semi-implicit method's smaller number of
matrix solutions per integration step.

Second, the HT99 network employs automated linking of reactions in
the data set to the species being evolved. This is in contrast to
the manual linking employed by WT90 and many older reaction
networks.  This automated linking helps to avoid implementation
mistakes, as we discovered while performing tests of NOVA in order
to understand the source of differences in the results of the
simulations between two versions of the code which used the same
reaction rate library but different nuclear reaction networks.  In
these tests, we discovered that while the REACLIB dataset used in
our prior studies (S98, S00), included the {\it pep} reaction ($p
+ e^{-} +p \rightarrow d + \nu$), it was not linked to abundance
changes in the WT90 network. While for Solar modeling energy
generation from the {\it pep} reaction is unimportant (but not the
neutrino losses), in the WD envelope the density can reach to
values of $10^4$ gm cm$^{-3}$ which is about two orders of
magnitude larger than in the core of the Sun. The increased
density increases the rate of energy generation by about 40\% over
calculations with the $pep$ reaction absent. The increased energy
generation then has the effect of reducing the amount of accreted
material since the temperature rises faster per gram of accreted
material. (The effect of changes in the rate of energy generation
on simulations of the CN outburst is discussed in detail in S98.)
Given a smaller amount of accreted material at the time when the
steep temperature rise begins in the TNR, the nuclear burning
region is less degenerate and, therefore, the peak temperatures
are lower for models evolved with the same nuclear reaction rate
library used in our previous studies (see below).

Finally, we use the analytic fitting formulas of Itoh et al.
(1996) for the neutrino energy loss rates from pair ($e^+ + e^-
\rightarrow \nu_e + {\bar \nu}_e$), photo ($e^{\pm} + \gamma
\rightarrow e^{\pm} + \nu + \bar{ \nu}_e$), plasma ($\gamma_{{\rm
plasmon}} \rightarrow \nu_e + \bar{ \nu}_e$), bremsstrahlung ($e^-
+ A^Z \rightarrow  e^- + A^Z + \nu_e + \bar{\nu} _e$), and
recombination ($e^-_{\rm continuum} \rightarrow e^-_{\rm bound} +
\nu_e + \bar{ \nu}_e$) processes. As stellar evolution codes
generally require derivative information for the Jacobian matrix,
our implementation of the Itoh et al. (1996) fitting formulas
(available from \url{http://www.cococubed.com}) returns the
neutrino loss rate and its first derivatives with respect to
temperature, density, $\bar{A}$ (average atomic weight) and
$\bar{Z}$ (average charge).

\section{Evolutionary Sequences using Four Nuclear Reaction Libraries}
\subsection{The Initial Models and Libraries}

Our calculations were done with 95 zone, 1.35M$_\odot$ complete
WDs. As in S98 and S00, we assume that the material being accreted
from the donor star is of Solar composition but that it has
already mixed with the core material so that the actual accreting
composition chosen for this study is 50\% Solar and 50\% ONeMg
material. We assume a value of 2 for the mixing-length to scale
height ratio ($l/H_p$). All other details of our calculations
(opacities, equations of state, etc.) are described in S98 and
S00.

We evolved four different sequences using a different reaction
rate library for each sequence but the same nuclear reaction
network (HT99).  The reaction rate library used in Politano et al.
(1995) included the rates from Caughlan and Fowler (1988) and
Thielemann et al. (1987, 1988).  They were compiled by Thielemann
and made available to Truran and Starrfield and also used for the
calculations reported in WT90 (P1995 in both plots and tables).
S98 used an updated reaction rate library which contained new
rates calculated, measured, and compiled by Thielemann and
Wiescher (labeled S98 in both plots and tables). A discussion of
the improvements is provided in S98.  The third library is
described in Iliadis et al. (2001) and was used for the
simulations in Starrfield et al. (2001). It is labeled I2001. The
last library used in ``This Work'' is the August 2005 compilation
of Iliadis. This library is an updated version of the library
described in Iliadis et al. (2001) and used in Starrfield et al.
(2001). A detailed description of this library will appear in
Starrfield et al. (2006, in prep.).  There is one additional
calculation in Table 1. A comparison calculation done with the
Politano et al. (1995) reaction rate library and the WT90 nuclear
reaction network. As noted above, this network does not include
the $pep$ reaction and we provide it here only for comparison. We
have recalculated the simulation for this paper using the same
equations of state and opacities as used for the other
calculations.

\subsection{The Evolutionary Results}

The initial properties of the WD are provided in the table
comments.  We evolved five evolutionary sequences. In all cases,
we assumed an initial WD luminosity of $\sim 4 \times
10^{-3}$L$_\odot$ and a mass accretion rate of 10$^{16}$ gm
s$^{-1}$ ($1.6 \times 10^{-10}$M$_\odot$ yr$^{-1}$).  This mass
accretion rate is 5 times lower than the lowest rate used in
either S98 or S00 and was chosen to maximize the amount of
accreted material given the increased energy generation from the
$pep$ reaction.  Numerous studies of accretion onto WDs by many
different authors demonstrate that the results of the evolution
depend strongly on the initial WD luminosity and mass accretion
rate (c.f., Yaron et al. 2005, and references therein).

\begin{table}
\begin{tabular}{lccccc}
\hline Reaction Library:\tablenote{The initial model for all 5
evolutionary sequences had M$_{\rm WD}$=1.35M$_\odot$, L$_{\rm
WD}$=$4.2 \times 10^{-3}$L$_\odot$, T$_{\rm eff}$=$2.5 \times
10^4$K, R$_{\rm WD}$=2495 km, and a central temperature of $1.2
\times 10^7$K}&P1995\tablenote{Library used in Politano et al.
(1995): pep reaction not included(WT90
network)}&P1995\tablenote{Library used in Politano et al. (1995):
pep reaction included (HT99 network)} &S1998\tablenote{Library
used in Starrfield et al. (1998): pep reaction included (HT99
network)}&I2001\tablenote{Library described in Iliadis et al.
(2001): pep reaction included (HT99 network)} &This
Work\tablenote{Iliadis (2005: this work) library: pep reaction
included (HT99 network)}\\
\hline
$\tau$$_{\rm acc}$($10^5$ yr)&2.5&2.1&2.1&2.1&1.8\\
M$_{\rm acc}$(10$^{-5}$M$_{\odot}$)&3.9& 3.3& 3.3 &  3.3 & 2.8 \\
T$_{\rm peak}$($10^6$K)&459& 413  & 414 &  407& 392 \\
$\epsilon_{\rm nuc-peak}$(10$^{17}$erg gm$^{-1}$s$^{-1}$)&22.8 & 8.4& 8.6  & 4.9  & 4.4 \\
L$_{\rm peak}$ (10$^5$L$_{\odot}$)&8.0&  9.6& 8.0 &  7.3 & 5.9 \\
T$_{\rm eff-peak}$($10^5$K)&20&13&13&8.8&8.8 \\
M$_{\rm ej}$(10$^{-5}$M$_{\odot}$)&3.3& 2.3& 2.3 &  2.3 & 1.7\\
V$_{\rm max}$(km s$^{-1}$)&6050 &5239& 4755 & 4787 & 4513 \\
\hline
\end{tabular}
\caption{Initial Parameters and Evolutionary Results}
\end{table}

We use the same composition for the accreting material as in
Politano et al. (1995; see also: S98; S00; and Starrfield et al.
2001: a mixture of half-solar and half-ONeMg by mass fraction). By
using this composition, we assume that core material has mixed
with accreted material from the beginning of the evolution. This
composition also effects the amount of accreted mass at the peak
of the TNR since it has a higher opacity than if no mixing were
assumed.  The results of our evolutionary calculations are given
in Table 1 and 2.  Table 1 gives the initial parameters and
evolutionary results and Table 2 gives the abundances of the
ejected material (by mass) for the 4 different simulations.  We do
not report the abundance results for the sequence done without the
$pep$ reaction since this calculation is not realistic.  The
evolutionary parameters are provided only to demonstrate the
effects of including this reaction on the evolution.

The rows in Table 1 are the reaction rate library, the accretion
time to the TNR($\tau$$_{\rm acc}$), the accreted mass(M$_{\rm
acc}$), peak temperature in the TNR (T$_{peak}$), peak rate of
energy generation during the TNR, ($\epsilon_{\rm nuc-peak}$),
peak luminosity (L$_{\rm peak}$), peak effective temperature
(T$_{\rm eff-peak}$), ejected mass (M$_{\rm ej}$), and the peak
expansion velocity after the radii of the surface layers have
reached $\sim 10^{13}$cm (V$_{\rm max}$). By this time the outer
layers are optically thin, have far exceeded the escape velocity,
and there is no doubt that they are escaping.

As noted above, we provide two different columns for Politano et
al. (1995). The first, with the superscript \dag, is taken from a
calculation done for this paper using the Politano et al. library
and the WT90 network in NOVA. The second, with the superscript
$**$, uses the same reaction rate library as Politano et al. but
the energy generation and nucleosynthesis is obtained with the
HT99 network. These two columns, therefore, show the effects of
including the $pep$ reaction on the TNR simulations.

Table 1 shows that the largest change in the results of the
evolution occurs with the inclusion of the $pep$ reaction.  If we
compare P1995$^\dag$ and P1995$^{**}$, then the $\sim$40\%
increase in energy production from just adding the $pep$ reaction
to the network results in a decrease of $\sim$19\% in both
accretion time and accreted mass. The large change must be caused
by the addition of the $pep$ reaction since the two reaction rate
libraries are otherwise identical.  Because there is less accreted
mass on the WD at the time of the TNR (comparing the sequences
done with the $pep$ reaction included to the one without it), the
peak temperatures do not reach to as high values as in the
sequence with the $pep$ reaction absent.  If we compare the
accretion time and accreted mass for the four sequences with the
$pep$ reaction included, we see that changes in the nuclear
reaction rate library have a sizable effect.  In addition, there
have been some small changes to the reaction rates in the
proton-proton chain and that is where the sequences spend most of
their time during the accretion phase.

\begin{figure}
\includegraphics[height=0.35\textheight]{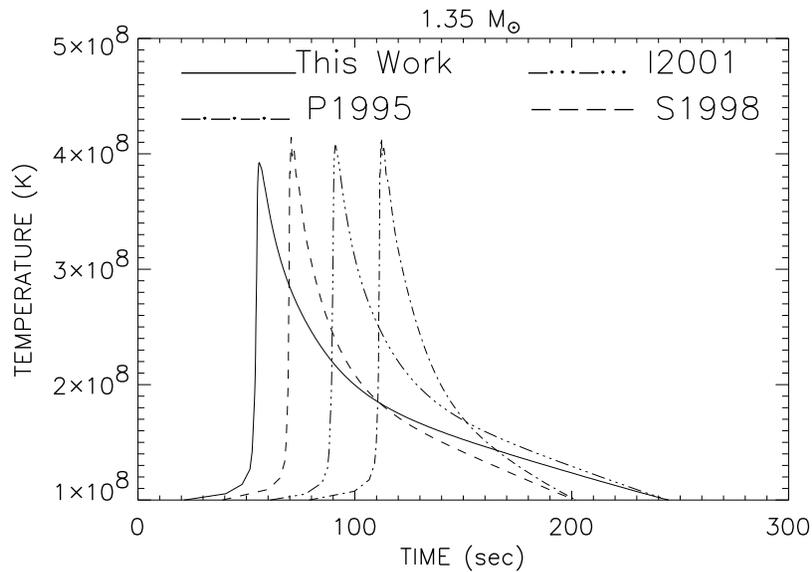}
\centering \caption{The variation with time of the temperature in
the deepest hydrogen-rich zone around the time when peak
temperature occurs.  We have plotted the results for four
different simulations on a 1.35M$_\odot$ WD. The identification
with calculations done with a specific library is given on the
plot. In this plot and all following plots, S1998 refers to
Starrfield et al. (1998:S98), P1995 refers to Politano et al.
(1995), I2001 refers to Iliadis et al. (2001), and This Work
refers to this paper. The details of the associated reaction rate
library are given in the text.  The curve for each sequence has
been shifted slightly in time to improve its visibility.}
\end{figure}

Figure 1 shows the variation of temperature with time for the
deepest hydrogen-rich zone and we plot only the simulations done
with the $pep$ reaction included.  The specific evolutionary
sequence is identified on the plot.  The reference to the reaction
network used for that calculation is given in the caption for
Figure 1. The time coordinate is arbitrary and chosen to clearly
show each curve.  This figure shows that there are important
differences between the four simulations. Peak temperature drops
from about 413 million degrees to 392 million degrees and peak
nuclear energy generation drops by about a factor of 2.5 from the
oldest library to the newest library ($8.4 \times 10^{17}$erg
gm$^{-1}$s$^{-1}$ to $4.4 \times 10^{17}$erg gm$^{-1}$s$^{-1}$).
The temperature declines more rapidly for the sequence computed
with the oldest reaction library (Politano et al. 1995) because it
exhibited a larger release of nuclear energy throughout the
evolution which caused the overlying zones to expand more rapidly
and the nuclear burning layers to cool more rapidly. In contrast,
the newest library, with the smallest expansion velocities, cools
slowly.

\begin{table}
\begin{tabular}{@{}lcccc}
\hline Reaction Library:&P1995\tablenote{Library used in Politano
et al. (1995): pep reaction included (HT99 network)}
&S1998\tablenote{Library used in Starrfield et al. (1998): pep
reaction included (HT99 network)}&I2001\tablenote{Library
described in Iliadis et al. (2001): pep reaction included (HT99
network)} &This Work\tablenote{Iliadis (2005: this work) library:
pep reaction included (HT99 network)}\\
\hline

H &0.27 &0.27 &0.27&0.28 \\
$^4$He&0.18&0.18&0.17& 0.17\\
$^{12}$C  &$8.0 \times 10^{-3}$&$1.2 \times 10^{-2}$& $8.0 \times 10^{-3}$&$6.2 \times 10^{-3}$ \\
$^{13}$C  &$2.8 \times 10^{-3}$&$4.0 \times 10^{-3}$& $2.4 \times 10^{-3}$&$2.4 \times 10^{-3}$ \\
$^{14}$N  &$4.3 \times 10^{-3}$&$4.8 \times 10^{-3}$& $4.3 \times 10^{-3}$&$8.4 \times 10^{-3}$ \\
$^{15}$N  &$0.11$&0.11& $6.8 \times 10^{-2}$&$6.0 \times 10^{-2}$ \\
$^{16}$O  &$1.2 \times 10^{-3}$&$1.1 \times 10^{-3}$& $2.4 \times 10^{-3}$&$2.4 \times 10^{-3}$ \\
$^{17}$O  &$1.1 \times 10^{-3}$&$1.0 \times 10^{-3}$& $5.9 \times 10^{-2}$&$6.7 \times 10^{-2}$ \\
$^{18}$O  &$7.8 \times 10^{-3}$&$6.7 \times 10^{-3}$& $3.0 \times 10^{-3}$&$1.5 \times 10^{-3}$ \\
$^{18}$F  &$2.5 \times 10^{-3}$&$2.3 \times 10^{-3}$& $9.3 \times 10^{-4}$&$5.9 \times 10^{-4}$ \\
$^{22}$Na &$3.5 \times 10^{-2}$&$5.1 \times 10^{-2}$& $3.0 \times 10^{-2}$&$2.3 \times 10^{-2}$ \\
$^{24}$Mg &$2.8 \times 10^{-3}$&$1.9 \times 10^{-3}$& $2.1 \times 10^{-3}$&$1.9 \times 10^{-3}$ \\
$^{26}$Mg &$1.6 \times 10^{-2}$&$1.2 \times 10^{-2}$& $2.2 \times 10^{-3}$&$1.5 \times 10^{-3}$ \\
$^{26}$Al &$2.8 \times 10^{-3}$&$2.1 \times 10^{-3}$& $2.7 \times 10^{-3}$&$3.0 \times 10^{-3}$ \\
$^{27}$Al &$2.8 \times 10^{-2}$&$3.4 \times 10^{-2}$& $1.4 \times 10^{-2}$&$1.4 \times 10^{-2}$ \\
$^{28}$Si &$2.3 \times 10^{-2}$&$3.5 \times 10^{-2}$& $2.7 \times 10^{-2}$&$2.9 \times 10^{-2}$ \\
$^{29}$Si &$6.3 \times 10^{-3}$&$7.0 \times 10^{-3}$& $1.9 \times 10^{-2}$&$1.8 \times 10^{-2}$ \\
$^{30}$Si &$2.3 \times 10^{-2}$&$2.4 \times 10^{-2}$& $3.1 \times 10^{-2}$&$3.8 \times 10^{-2}$ \\
$^{31}$P &$3.0 \times 10^{-2}$&$3.0 \times 10^{-2}$& $4.3 \times 10^{-2}$&$3.7 \times 10^{-2}$ \\
$^{32}$S  &$2.1 \times 10^{-2}$&$2.8 \times 10^{-2}$& $3.9 \times 10^{-2}$&$4.0 \times 10^{-2}$ \\
$^{34}$S  &$1.5 \times 10^{-3}$&$1.3 \times 10^{-3}$& $1.3 \times 10^{-3}$&$7.2 \times 10^{-4}$ \\
$^{36}$Ar  &$6.1 \times 10^{-4}$&$2.2 \times 10^{-4}$& $1.5 \times 10^{-4}$&$6.8 \times 10^{-5}$ \\
$^{40}$Ca  &$2.1 \times 10^{-5}$&$2.4 \times 10^{-5}$& $2.4 \times 10^{-5}$&$1.8 \times 10^{-5}$ \\
\hline
\end{tabular}
\caption{Comparison of the Ejecta Abundances for 1.35M$_\odot$
White Dwarfs (All abundances are mass fraction)}
\end{table}

The differences in total nuclear energy generation (L/L$_\odot$)
as a function of time for each mass is shown in Figure 2.  The
time coordinate is consistent with that used in Figure 1. Note
that peak nuclear energy production for the latest library is
definitely lower than seen in the earlier libraries. The changes
in the libraries are more important for the more massive isotopes
and become more important as higher temperatures are reached.

The abundance predictions, for the ejected material, from our four
evolutionary sequences are given as mass fraction in Table 2. Here
we discuss only the most important nuclei.  The increase in the
$^4$He abundance is small compared to the observed helium
abundances in CN ejecta which can reach, if not exceed, 0.5 (G98).
The large helium abundance, in combination with the large observed
CNO abundances, is strong evidence for mixing of the accreted
material with layers in the WD underlying the accreting material
at some time during the outburst. The large helium abundances
observed in RN such as U Sco or V394 CrA suggest that mixing has
also occurred in these systems even if their total CNO abundances
are not dramatically enriched over solar.  Examining the behavior
of the individual abundances, we see that $^{12}$C and $^{13}$C
are virtually unchanged by the updated reaction rates. In
contrast, the abundance of $^{14}$N nearly doubles and that of
$^{15}$N decreases by a factor of two going from the first to the
latest reaction rate library.  $^{16}$O also doubles in abundance
while $^{17}$O grows by a factor of 60 and becomes the most
abundant of the CNO nuclei in the ejecta.  For this WD mass, the
C/O ratio is 0.12.

\begin{figure}
\includegraphics[height=0.35\textheight]{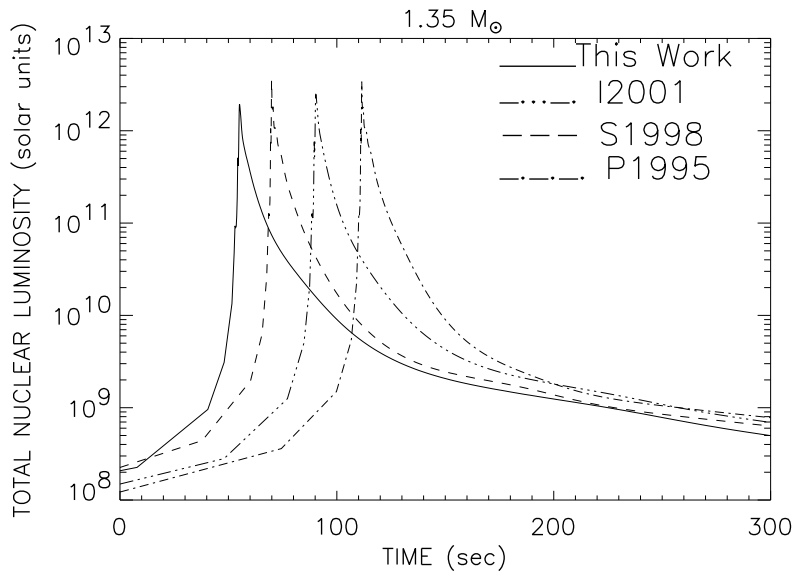}
\centering \caption{The variation with time of the total nuclear
luminosity (erg s$^{-1}$) in solar units (L$_\odot$) around the
time of peak temperature during the TNR on a 1.35M$_\odot$ WD. We
integrated over all zones taking part in the explosion. The
identification with each library is given on the plot. The time
coordinate is chosen to improve visibility. }
\end{figure}

The abundance of $^{22}$Na decreases with the library update and
$^{24}$Mg is severely depleted by the TNR.  $^{26}$Al is unchanged
by the changes in the reaction rates while the abundance of
$^{27}$Al drops by a factor of two.  This result implies that TNRs
on more massive WDs eject about the same fraction of $^{26}$Al as
$^{27}$Al.  We also find that contrary to Politano et al. (1995)
that the amount of $^{26}$Al ejected is virtually independent of
WD mass.  All the Si isotopes ($^{28}$Si, $^{29}$Si, and
$^{30}$Si) are enriched in the calculations done with the latest
library. Finally, while the ejecta abundances of $^{36}$Ar, and
$^{40}$Ca have declined as the reaction rate library has been
improved, they are all produced in the nova TNR since their final
abundances exceed the initial abundances.

\section{Summary}

In this paper we examined the consequences of improving the
nuclear reaction library on our simulations of TNRs on
1.35M$_\odot$ WDs.  We have found that the changes in the rates
have affected the nucleosynthesis predictions of our calculations
but not, to any great extent, the gross features of the evolution.
In addition, we have used a lower mass accretion rate than in our
previous studies in order to accrete (and eject) more material.
This has, as expected, caused the peak values of some important
parameters to increase over our previous studies at the same WD
mass. However, because some important reaction rates have declined
in the new reaction rate library this has not increased the
abundances for nuclei above aluminum and, in fact, they have
declined while the abundances of both $^{26}$Al and $^{27}$Al have
increased. In contrast, the abundance of $^{22}$Na has declined
from the values predicted in our earlier work.

\begin{theacknowledgments}

S. Starrfield acknowledges partial support from NSF and NASA
grants to ASU. He also thanks J. Aufdenberg and ORNL for generous
allotments of computer time. WRH is partly supported by the NSF
under contracts PHY-0244783 and and by the DOE, through the
Scientific Discovery through Advanced Computing Program. ORNL is
managed by UT-Battelle, LLC, for the U.S. DOE under contract
DE-AC05-00OR22725.

\end{theacknowledgments}

\end{document}